\documentclass[11pt,twoside]{article}
\usepackage{newpasp}
\usepackage{graphicx}

\def\edcomment#1{\iffalse\marginpar{\raggedright\sl#1\/}\else\relax\fi}
\reversemarginpar

\begin{document}
\title{A first look at cataclysmic variable stars from the 2dF
QSO survey} 

\author{T.R.Marsh, L.Morales-Rueda, D.Steeghs}  

\affil{Dept of Physics and Astronomy,
University of Southampton, Highfield, Southampton SO17~1BJ,
UK}

\author{P.Maxted}
 
\affil{Dept of Physics and Astronomy,
Keele University, Keele, Staffs, UK}

\author{U.Kolb} 

\affil{Dept of Physics and Astronomy, Open University, 
Milton Keynes, UK}

\author{Brian Boyle, Scott Croom}

\affil{Anglo-Australian Observatory, PO Box 296, Epping, NSW 2121,
Australia}

\author{Nicola Loaring, Lance Miller}

\affil{Department of Physics, Oxford University, Keble Road, Oxford
OX1~3RH, UK}

\author{Phil Outram, Tom Shanks}

\affil{Physics Department, University of Durham, South Road, Durham
DH1~3LE, UK}

\author{Robert Smith}

\affil{
Astrophysics Research Institute, Liverpool John Moores University, Twelve
Quays House, Egerton Wharf, Birkenhead, CH41 1LD, UK}

\begin{abstract}
The 2dF QSO survey is a spectroscopic survey of $48$,$000$
point-sources selected by colour with magnitudes in the range $18.35
\la B \la 20.95$.  Amongst QSOs, white dwarfs, narrow-line galaxies
and other objects are some cataclysmic variables. This survey should
be sensitive to intrinsically faint CVs. In the standard picture of CV
evolution, these form the majority of the CV population.  We present
the spectra of 6 CVs from this survey. Four have the spectra of dwarf
novae and two are magnetic CVs. We present evidence that suggests that
the dwarf novae have period $P < 2\,{\rm h}$ and are indeed
intrinsically less luminous than average.  However, it is not clear
yet whether these systems are present in the large numbers predicted.
\end{abstract}

\section{Introduction}
The standard picture of cataclysmic variable (CV) evolution was
developed to explain three outstanding features of the orbital
period distribution of CVs: (1) the cutoff for $P  \ga 10\,{\rm h}$,
(2) the dearth of systems with $2 \la P \la 3\,{\rm h}$
(the ``period gap'') and (3) the cutoff for $P \la 80\,{\rm min}$. 
The theory involves magnetic braking at long periods,
gravitational radiation braking at short periods, and a hiatus in
between when the magnetic braking is disrupted resulting in the period
gap. Degeneracy of the donor stars at short periods leads to the period minimum.

This theory suffers from several problems. First, single star
studies provide no support for either the magnitude of magnetic
braking assumed in CV studies or its disruption when the donor star
becomes fully convective (Andronov, Pinsonneault, \& Sills 2001). Moreover, an
excess of systems is expected close to the period minimum, but is not
observed (Kolb \& Baraffe 1999; also Kolb, King, this
volume). Associated with this is a prediction that most (99\%) systems
should be of short period ($P < 2$hrs), whereas the observed number is
more like 50\% (Kolb 1993). This is also seen in the
observed space densities which are consistently lower than theoretical
predictions (see G\"{a}nsicke, this volume).

A possible reason for the absence of an excess of short period systems
is that they exist, but just have not been discovered. The only way
for this to be the case is if they never, or very rarely, go into
outburst. The uncertainty over accretion physics is large enough that
no one is competent to say whether or not this is possible, and so it
must at least be admitted as a possibility. If so, then we need to
search for such systems without relying on gross changes in
brightness. In this paper we present the first spectra of targets found
in a search based upon selection for ultraviolet excess. We start
by describing the survey and why we expect it to turn up 
intrinsically faint CVs. We then present the first spectra of the 6 
targets and discuss the evidence that suggests that they are indeed
of low luminosity. We finish by discussing whether these systems
can solve the problem of the missing submerged iceberg of the 
CV population outlined above.

\section{The 2dF QSO survey}
The 2dF (2-degree field) spectrograph on the $3.9$m Anglo-Australian
Telescope is a prime-focus, fibre-fed spectrograph that can obtain up
to 400 spectra at once covering a two-degree-wide field of the
sky. Two large projects have been running on this instrument since it
was commissioned: the 2dF Galaxy \& QSO surveys. The 2dF QSO survey
(Boyle et al. 2000; Croom et al. 2001),
which is still taking place (2001), will obtain spectra of $\approx
48$,$000$ sources which appeared star-like on archive blue photographic plates,
selected by colour and magnitude to exclude stars in the Milky
Way. The survey covers 740 sq.\ deg.\ in two strips, one near the
equator, and the other at $\delta \approx -30^\circ$ at the South
Galactic Pole.  CVs display ultraviolet excesses, thus they are
also included in the target list, and can be identified once
spectra are obtained. Such identification is in any case essential
because only about 60\%\ of the targets are expected to be QSOs.

\begin{figure}
\includegraphics[width=0.98\textwidth]{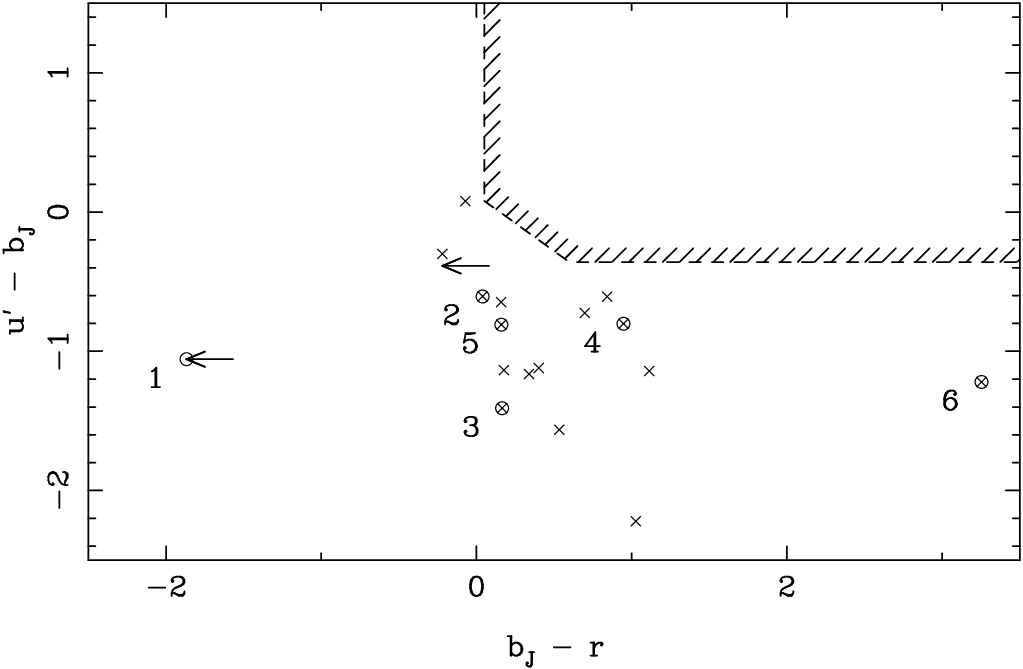}
\caption{Colour/colour plot of CV candidates from the 2dF QSO
survey. The region in the upper-right is occupied by normal stars and
is rejected. Most of the CV candidates are located well away from this
region, essentially because of their $u'-b_J$ colours. The numbers
along with the cross \& circle symbols correspond to the targets we
observed. Arrows indicate upper limits (no red detection). For our
targets, $u' \approx U - 0.24$, $b_J \approx B - 0.1$, and $r \approx
R$. The very red colour of target~6 indicates that it was in outburst
when the red plate was taken in March 1985 }
\label{fig:select}
\end{figure}

Fig.~\ref{fig:select} shows the locations of the CV candidates
in the colour selection plot used to select targets for the 2dF 
QSO survey. The plot does not indicate any build-up of candidates
towards the excluded zone, which would have suggested significant
incompleteness.

Colour-selected surveys are nothing new, some 30 CVs having
turned up in the PG survey (Green, Schmidt, \& Liebert 1986; Ringwald 1996).
However, the 2dF survey reaches much fainter apparent magnitudes than
did the PG survey. 2dF targets are selected to have $18.25 < b_J <
20.85$ ($\approx 0.1$ mags fainter still in $B$), so even the
brightest targets are fainter than the faintest PG CVs at $B \approx
16$. Of course, the area covered by the 2dF survey is smaller than the
PG survey's $10$,$700$ sq.\ deg., but it nevertheless has a larger
effective volume at faint absolute magnitudes.

\begin{figure}
\hspace*{\fill}
\includegraphics[width=0.98\textwidth]{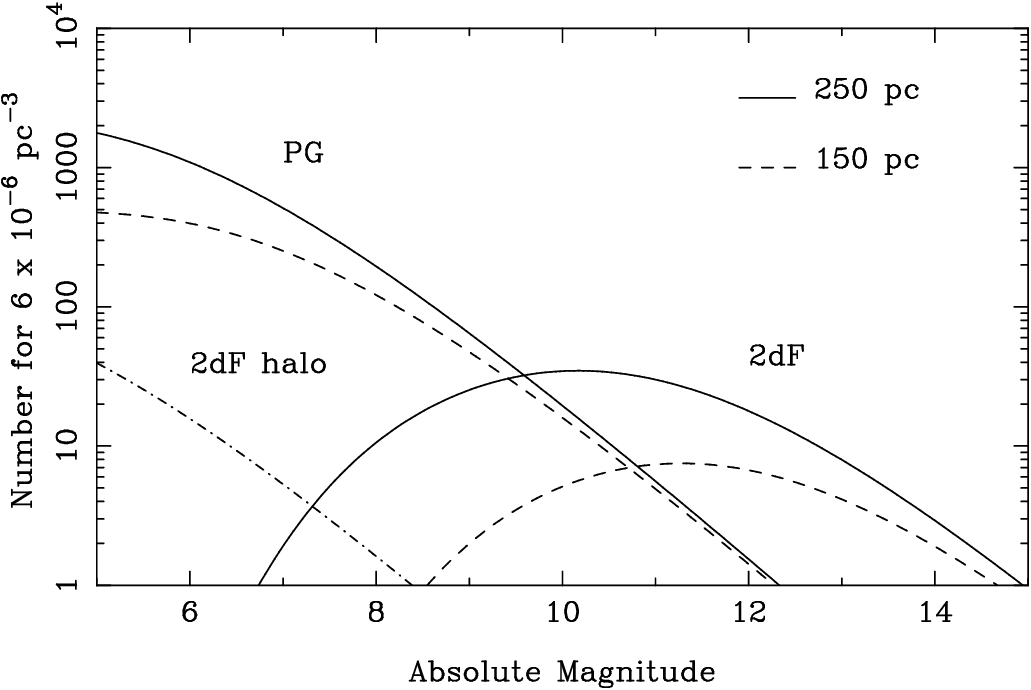}
\hspace*{\fill}
\caption{The number of systems detectable, assuming they all
have the same absolute magnitude, as a function of that magnitude
in the PG and 2dF surveys for two different scale heights in
the disk. We assume the local space density from Ringwald (1996) of
$6 \times 10^{-6}\,{\rm pc}^{-3}$. The dot-dash line in the lower
left shows the same for halo CVs assuming a space density $10^4$ times
smaller but an (arbitrary) $8\,{\rm kpc}$ scale height.}
\label{fig:volume}
\end{figure}

Assuming a particular absolute magnitude, galactic scale-height and
local space density, one can calculate the number of objects expected
in a given survey. We compare the effectiveness of PG and 2dF surveys
as a function of absolute magnitude in this manner in
Fig.~\ref{fig:volume}.  The PG survey is, as expected, more effective
at bright magnitudes because of its large area. The 2dF survey on the
other hand wins at absolute magnitudes $M > 9.5$ -- $10.5$, depending
upon the scale height assumed.  In principle the accretion luminosity
of CVs could descend to even fainter magnitudes than indicated on the
plot, however in such cases it is the magnitude of the white dwarf
that matters and those observed so far in CVs have always satisfied $M
< 13$.  For example, at $\log g = 8$, white dwarfs have absolute
magnitudes of $M_V = 12.1$ for a temperature $T = 10$,$000\,{\rm K}$
(Bergeron, Wesemael, \& Beauchamp 1995), corresponding to the coolest white dwarf
observed in any CV (Szkody, this conference).  Thus the 2dF
survey is sensitive to the faintest magnitudes known for any CV so
far. (The possibility of cooler white dwarfs should not be dismissed
however, because such systems may fail to appear in the 2dF and
similar UV excess surveys, thus the failure to find any cooler white
dwarfs in CVs may be a selection effect.)

\section{The Spectra}
As part of the survey classification, the 2dF spectra were checked by
eye. It was this that led to the identification of 19 CV candidates. 
We observed 6 of these
over three dark nights on the $4.2$m William Herschel Telescope (WHT)
in the Canary Islands in February 2001.  We used the ISIS spectrograph
to cover blue and red simultaneously.  On the blue arm we covered
$3460$ -- $5350$\AA\ at $3.5$\AA\ FWHM resolution, while on the red
arm we covered $6150$ -- $9116$\AA\ at $5.8$\AA\ resolution.

\begin{table}
\caption{Target details.}

\begin{tabular}{llllll}
\tableline
Target    & RA (J2000)         & Dec & $b_J$ & Type$^\S$ & Orbital period \\
\tableline
1  & 11 25 55.73&-00 16 38.9 & 19.64 & DN   & $0.0613$ d\\
2  & 12 10 05.30&-02 55 43.9 & 20.72 & DN   & --- \\
3  & 13 04 41.76&+01 03 30.8 & 20.73 & DN   & --- \\
4  & 14 22 56.32&-02 21 08.7 & 19.54 & Mag. & $0.1404$ d$^\dag$ \\
5  & 14 24 38.94&-02 27 39.9 & 19.52 & Mag. & $0.1555$ d$^\dag$ \\
6  & 14 35 00.18&-00 46 06.9 & 18.58 & DN   & $0.072727$ d$^\ddag$ \\
\tableline
\tableline
\end{tabular}\\
$\S$ Classified from spectra alone.\\
$\dag$ Several 1 cycle/day aliases possible as well.\\
$\ddag$ From Vanmunster, Velthuis, \& McCormick (2000).\\
\label{tab:targets}
\end{table}
\begin{figure}
\hspace*{\fill}
\includegraphics[width=0.98\textwidth]{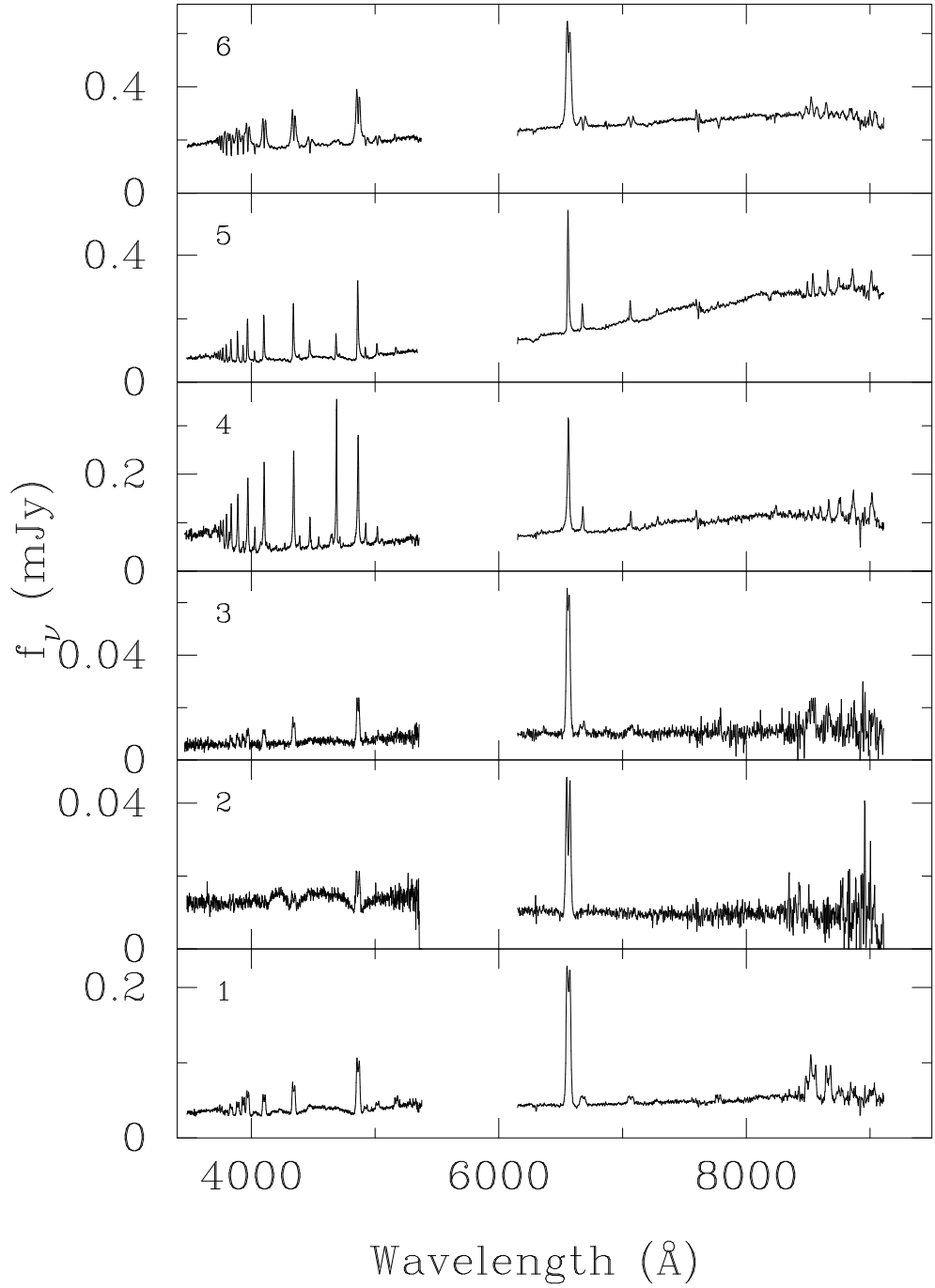}
\hspace*{\fill}
\caption{The average spectra of the six 2dF CV candidates observed.}
\label{fig:spectra}
\end{figure}
Details of the targets are shown in Table~\ref{tab:targets},
and their mean spectra are shown in Fig.~\ref{fig:spectra}
\footnote{A rather remarkable fact, given that only 100-odd magnetic
systems are known, is that the two magnetic systems found here are
within half a degree of each other!}.  As yet we have no photometry of
any of our targets beyond the target selection plates, and so we do
not know their outburst behaviour (except target~6 which has 
outbursts of $\approx 4$ mags, Vanmunster, Velthuis, \& McCormick 2000 and
Fig.~\ref{fig:select}).  However, the spectra are
characteristic enough that we can make identifications of the CV type
with confidence. Targets 1, 2, 3 and 6 have the spectra of quiescent
dwarf novae, while we identify targets~4 and 5 as ``magnetic'' (polar
or intermediate polar) because of their narrow lines and strong HeII
4686, and because the M-type donor star is visible in each of them
(each shows NaI 8200 and TiO bandheads at 7200 and 7500\AA). Novalike
variables can also show HeII 4686, but have broader lines and are
bright enough to drown out M-type donor stars. There is direct
evidence of cyclotron emission in target~5, visible in the red
spectrum in Fig.~\ref{fig:spectra} and in time-resolved spectra.

One of the aims of our observations was to measure orbital periods
(from variations in the emission lines) to test whether these systems
were indeed part of the predicted majority of short period systems. We
have been partly successful, and were helped by the discovery that
target~6 (our brightest, and also discovered in the Large Bright
Quasar Survey, Berg et al. 1992) was eclipsing
(Vanmunster, Velthuis, \& McCormick 2000).  We failed to find periods for
targets~2 and 3. We had only three spectra of target~3, while target~2
failed to show significant variability in the 10 spectra we acquired
(total of 4 hours exposure).  However, we suspect that, like targets 1
and 6, targets~2 and 3 are also short period ($P < 2$h), because there
are no obvious donor star features while at the same time the presence
of broad absorption wings shows that the white dwarf is visible.  In
addition, the lack of large radial velocity shifts in target~2, even
though its emission lines are clearly double-peaked (i.e. it is not of
low inclination), suggests that its donor is of low mass, as expected
at short orbital periods.

The short periods of the dwarf novae support the
theory, but given that 65 out of 121 dwarf novae with known period
have $P < 2.5\,{\rm h}$, no great significance can be attached to
this (Ritter \& Kolb 1998). Of rather more interest are the broad absorption
wings from the white dwarf in targets~1, 2 and 3, suggesting
$9 \la M_V \la 13$. In addition these
systems show steeper-than-typical Balmer decrements and clear double-peaked lines,
indicative of relatively cool disks of low optical depth.
\begin{figure}
\hspace*{\fill}
\includegraphics[width=0.98\textwidth]{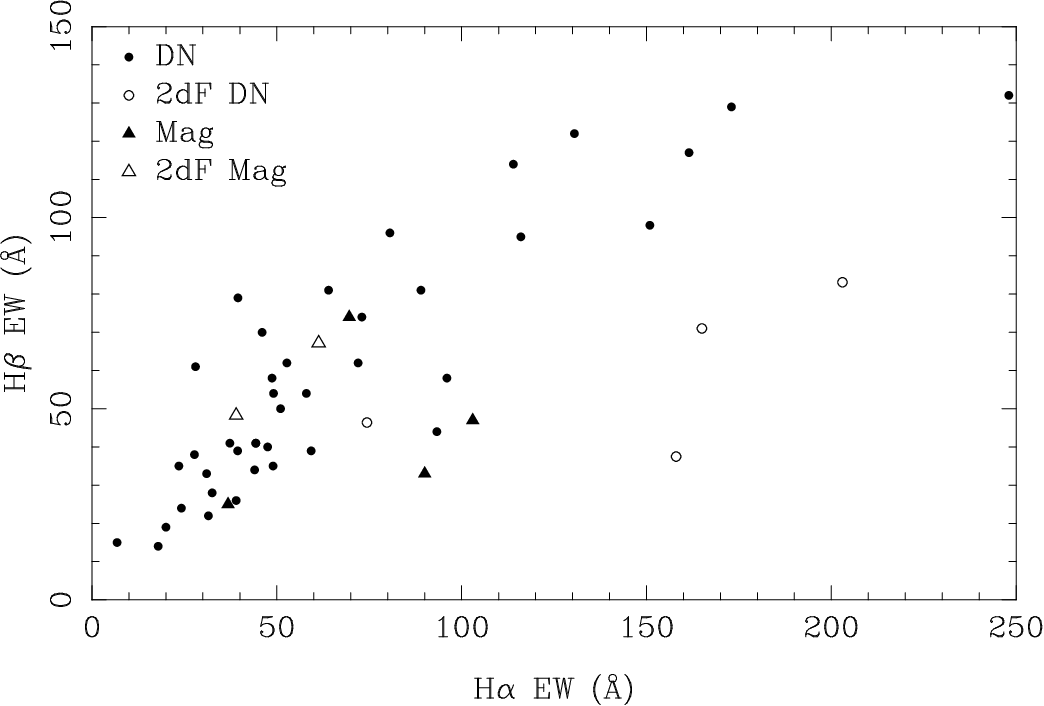}
\hspace*{\fill}
\caption{H$\alpha$ and H$\beta$ equivalent widths of the 2dF CVs
compared with the compilation of Echevarria (1988).}
\label{fig:balmer}
\end{figure}
The H$\alpha$ and H$\beta$ equivalent widths are compared in Fig.~\ref{fig:balmer} with
those listed in the compilation of Echevarria (1988).
Targets~1, 2 and 3 lie in a region of fairly average H$\beta$ flux,
but strong H$\alpha$, not occupied by the systems in Echevarria's
list (although there are stars such as GD552 and LL~And that would
probably be in a similar position).

\section{Discussion}
The 2dF CVs show signs of being of low intrinsic luminosity, but it is
too early to say whether they match expectations in all
respects. Certainly target~1, with a period of $88\,{\rm min}$ could
be a ``post period-bounce'' system, but precise periods are required
for targets~2 and 3 as well. An interesting point to note about these
systems is that it may well be the white dwarfs that dominate rather
than accretion light. Thus searches for ``flickering'' may prove
disappointing as a means of detecting low-$L$ systems. On the other
hand a subset of them should show brief and deep eclipses, so
variability searches are still worthwhile, if sufficient area can be
covered.

The 2dF survey is about 2/3$^{\rm rd}$ complete.  At the moment there
are about $20$ CV candidates. Not all of these will be confirmed as CVs
perhaps, but others may be found, and so a total of 30
seems reasonable. This is consistent with the PG survey for
Ringwald's (1996) space density, a scale-height of $250\,{\rm
pc}$, and a CV luminosity function dominated by systems with $M_B
\approx 10$. A lower scale height would require a higher space density
(see Fig.~\ref{fig:volume}) approaching the region implied by theory.
At this stage one might say that the uncertainties are such that there
is little evidence for any discrepancy between observations and
theory. However, the key test will be provided by their
periods, because we expect the ``missing majority'', if they
exist, to cluster close to the period minimum. Confirming or refuting
this is our long term aim.

\section{Conclusions}
We have presented spectra of 6 cataclysmic variables discovered in the
2dF QSO survey. Four are dwarf novae, two are magnetic. Three of the dwarf
novae show white dwarf absorption wings, steep Balmer decrements and
double-peaked line emission, with no sign of their donor stars. All
are characteristics of low-luminosity, short-period systems. One of
the three has a confirmed period of $88\,{\rm mins}$, but further
measurements of orbital periods are needed to see if these systems
cluster near the orbital period minimum. The fourth and brightest
dwarf nova has a period $105\,{\rm mins}$; it is an eclipsing system
very similar to OY~Car and Z~Cha. We point out that very low
luminosity CVs are likely to be dominated by light from their white
dwarfs rather than their accretion disks or stream/disk impact regions.


\begin{references}
\reference Andronov, N., Pinsonneault, M., Sills, A. 2001, \apj, submitted, astroph/0104265

\reference Berg, C., Wegner, G., Foltz, C.~B., Chaffee, F.~H., Hewett, P.~C. 1992, \apjs, 78, 409

\reference Bergeron, P., Wesemael, F., Beauchamp, A. 1995, \pasp, 107, 1047

\reference Boyle, B.~J. et al. 2000, \mnras, 317, 1014

\reference Croom, S.~M. et al. 2001, \mnras, 322, L29

\reference Echevarria, J. 1988, \mnras, 233, 513

\reference Green, R.~F., Schmidt, M., Liebert, J. 1986, \apjs, 61, 305

\reference Kolb, U. 1993, \aap, 271, 149

\reference Kolb, U., Baraffe, I. 1999, \mnras, 309, 1034

\reference Ringwald, F.~A. 1996, ASSL Vol. 208: IAU Colloq. 158: Cataclysmic Variables and Related Objects

\reference Ritter, H., Kolb, U. 1998, \aaps, 129, 83

\reference Vanmunster, T., Velthuis, F., McCormick, J. 2000, Informational Bulletin on Variable Stars, 4955, 1

\end{references}
\end{document}